\documentclass{optica-article}

\journal{opticajournal} 

\articletype{Research Article}

\begin{document}

\title{Superposition- and interference-induced optical spectrum distortion in the figure-9 fiber laser}

\author{Xiang Zhang,\authormark{1,3} Yongzhuang Zhou,\authormark{1,3} Chengjie Gao,\authormark{1,3} Kangrui Chang,\authormark{1,3} Yong Shen,\authormark{1,3} Guochao Wang,\authormark{2,4} Hongxin Zou\authormark{1,3,5}}

\address{\authormark{1}Institute for Quantum Science and Technology, College of Science, National University of Defense Technology, Changsha 410073, China\\
\authormark{2}College of Intelligence Science and Technology, National University of Defense Technology, Changsha 410073, China\\
\authormark{3}Hunan Key Laboratory of Mechanism and Technology of Quantum Information, Changsha 410073, China\\
\authormark{4}wgc.19850414@163.com\\
\authormark{5}hxzou@nudt.edu.cn}



\begin{abstract*} 
The output pulse spectra of the figure-8 and figure-9 lasers typically exhibit more pronounced distortion than those from mode-locked lasers based on other saturable absorbers, as well as the spectra of their own intracavity pulses. 
Here, we demonstrate two figure-9 lasers with repetition rates of 190.6 MHz and 92.4 MHz and introduce a self-designed beam splitter that exhibits minimal spectral filtering into the fiber loop to output two interference-free pulses. 
By applying superposition and interference calculations to the experimental spectra of these two pulses, we obtained calculated spectra and found that their characteristics agree with the distortion features observed in the experimental spectra from the other two ports where superposition and interference occur.
Therefore, we conclude that the severe spectral distortion is caused by spectral superposition and interference, rather than the commonly believed nonlinear effects.
Furthermore, analysis based on the interference theory of the figure-9 laser reveals that the $p$-components of the two intracavity light beams usually interfere with non-equal intensity at the beam splitter where interference occurs, while the $s$-components always interferes with almost equal intensity. This mechanism results in a significant yet stable spectral difference between the intracavity and output pulses. Moreover, a change in the pump power can amplify the difference between the two $s$-components, thereby leading to the emergence of a minor peak at the optical spectrum center.
These findings provide new perspectives for simulating spectra that closely resemble experimental results and deepen our understanding of spectral evolution and pulse dynamics of the figure-9 lasers.

\end{abstract*}

\section{Introduction}
Over the past twenty-five years, ultrashort pulse lasers have enabled numerous applications, including precision ranging\cite{trocha_ultrafast_2018}, time and frequency transfer\cite{gozzard_ultrastable_2022}, materials processing\cite{sugioka_will_2021}, attosecond science\cite{pupeza_extreme-ultraviolet_2021}, trace gas sensing\cite{link_dual-comb_2017}, and ultra-low noise microwave generation\cite{nakamura_coherent_2020}. 
The generation of ultrashort pulses through passive mode-locking typically relies on saturable absorbers. These can be real materials, including semiconductor saturable absorber mirrors (SESAMs) \cite{keller_coupled-cavity_1990}, carbon nanotubes, and graphene; or artificial equivalents, such as nonlinear polarization rotation (NPR) \cite{inaba_long-term_2006}, the nonlinear optical loop mirror (NOLM) \cite{doran_nonlinear-optical_1988}, and the nonlinear amplifying loop mirror (NALM) \cite{kuse_all_2016, hansel_all_2017}.
Mode-locked lasers based on the NOLM or NALM are categorized by their configuration as either figure-8 or figure-9 lasers. The figure-9 laser was invented in response to the poor self-starting ability and low repetition rate issues of the figure-8 lasers\cite{hansel_all_2017}. This new laser makes it easier to achieve mode-locking self-starting with a shorter cavity. The figure-9 lasers offer the combined advantages of all-polarization-maintaining (all-PM) fiber optics, easy mode-locking self-starting, and low cost. Hence, they are among the most popular choices at present. However, compared to the pulses from mode-locked lasers based on other saturable absorbers, the output pulses from the figure-9 laser exhibit severe spectral distortion, with the degree of distortion varying across different output ports.

Spectral distortions are commonly attributed to nonlinear effects and gain narrowing\cite{agrawal_nonlinear_2019,nishizawa_investigation_2019,laszczych_dispersion_2021,papadopoulos_compensation_2009}. The nonlinear effects include self-phase modulation (SPM), cross-phase modulation (XPM), self-steepening, and other higher-order nonlinear effects. The SPM effect plays a role in symmetric spectral broadening and narrowing~\cite{agrawal_self-phase_1989,washburn_transform-limited_2000}, while the intensity dependence of the group velocity leads to the self-steepening effect, which renders the SPM-induced broadening asymmetric\cite{moses_self-steepening_2007,nishizawa_investigation_2019,laszczych_dispersion_2021}. The XPM effect can also cause asymmetric spectral broadening, and its function is non-reciprocal~\cite{agrawal_nonlinear_2019,hsieh_cross-phase_2007}. At relatively high pulse peak powers, higher-order nonlinear effects become significant, leading to complex spectral changes. At conventional peak power levels, the spectral profiles of the mode-locked fiber lasers obtained through experiments and theoretical calculations are relatively regular~\cite{hao_study_2024,hu_generation_2023,zhang_route_2022,cheng_high-repetition-rate_2018}. This, however, stands in sharp contrast to the complex distortions observed in figure-9 lasers.
 
When the cavity length is very short or the net intracavity dispersion is very small, the spectrum of the figure-9 laser shows a relatively symmetric and regular pattern on a logarithmic intensity scale~\cite{mayer_flexible_2020,wanli_gao_446_2018}. 
As the cavity length increases, leading to higher net intracavity dispersion and a lower repetition rate, the spectral distortion gradually becomes severe\cite{han_environment-stable_2022,liu_all_2020,yang_all-polarization-maintaining_2022}. 
The worsening of spectral distortion is typically attributed to nonlinear effects, NALM asymmetry or interference effects~\cite{nishizawa_investigation_2019}. Currently, there is a lack of research addressing which of these is the actual mechanism.
Unlike in other mode-locked fiber lasers, the saturable absorption effects in the figure-8 and figure-9 lasers arise from the interference of two beams, and its amplitude noise is related to transmittance~\cite{edelmann_intrinsic_2021}.
If we investigate these unique characteristics and add more output ports to measure the spectra, it will help identify the spectral evolution and the formation mechanism of the spectral distortion. 

In this paper, we construct two all-PM figure-9 fiber lasers using a self-designed component that combines a tap coupler with wavelength division multiplexing (TCWDM). The TCWDM functions as a beam splitter, providing two additional output ports to separately extract the clockwise and counterclockwise pulses propagating in the fiber loop. This configuration allows us to obtain the spectra of the pulses without interference and to avoid using standard fiber couplers that introduce spectral filtering in the fiber loop. We performed numerical analyses based on interference and superposition theory on the experimental spectra of the two intracavity beams, and found that the characteristics of the calculated spectra are consistent with the spectral distortion observed at the other output ports.
Furthermore, through theoretical analysis of the saturable absorption effect in the figure-9 laser, it is found that the interference and superposition of the intracavity beams lead to severe spectral distortion of the output pulses, and the intensity difference between the intracavity components of the two interfering pulses is significantly different from that between the output components.

\section{Experiment setup}

\begin{figure}[htbp]
	\centering
	\includegraphics[width=0.8\textwidth]{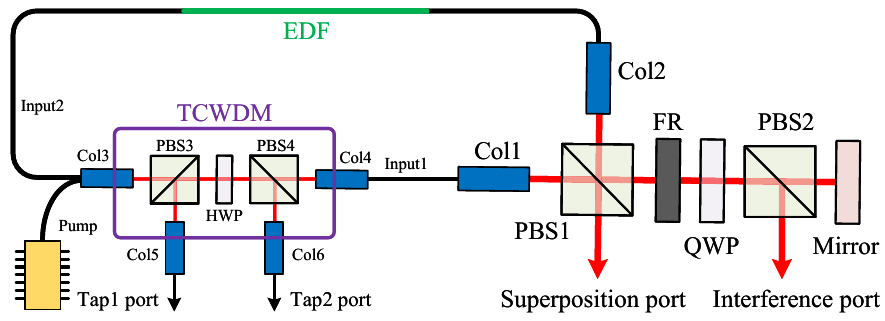}
	\caption{Experimental setup of the mode-locked fiber laser. Col1, Col2, Col4, Col5, Col6: collimators; Col3: WDM collimator; EDF: erbium-doped fiber; TCWDM: tap coupler hybrid with WDM; PBS1, PBS2, PBS3, PBS4: polarization beam splitters; QWP: quarter-wave plate; HWP: half-wave plate; FR: Faraday rotator. The four output ports are Tap1 port, Tap2 port, superposition port and interference port.}
	\label{fig1}
\end{figure}

The schematic of the all-PM figure-9 fiber laser is shown in Fig.~\ref{fig1}. The working distances of the two collimators (Col1 and Col2) are 80~mm. The polarizations of the slow-axis pulses output from Col1 and Col2 are parallel to the $p$ and $s$ states of the PBS1, respectively. These two orthogonally polarized light beams interfere on the PBS2 after passing through a non-reciprocal phase shifter composed of a Faraday rotator (FR) and a quarter-wave plate (QWP), showing a saturable absorption effect. The fiber loop consists of the passive fiber (PM1550) and a 392~mm erbium-doped fiber (PM-ESF-7/125, Nufern). The laser has four output ports, labeled Tap1 port, Tap2 port, superposition port and interference port. The Tap1 and Tap2 ports output pulses that have not undergone interference.

The tap ratio of the TCWDM is about 10\% and the pump light in the WDM collimator (Col3) is reflected to another fiber. The package length of the TCWDM is only 64~mm, which is shorter than that of the fiber device made by a fiber coupler and a WDM. In the TCWDM, the polarization of the slow-axis light from Col3 is parallel to the $p$ state of the PBS3, and that from Col4 is parallel to the $p$ state of the PBS4. 
After passing through PBS3 and a half-wave plate (HWP), the light from Col3 experiences polarization rotation and is split into two beams at PBS4, which are coupled into Col4 and Col6, respectively.
Similarly, the light from Col4 passes through PBS4, undergoes polarization rotation via the HWP, and is then split into two beams at PBS3, which are coupled into Col3 and Col5, respectively. The angle of the HWP determines the tap ratio of the TCWDM.

\section{Results and Discussion}
\begin{figure}[htbp]
	\centering
	\includegraphics[width=0.85\textwidth]{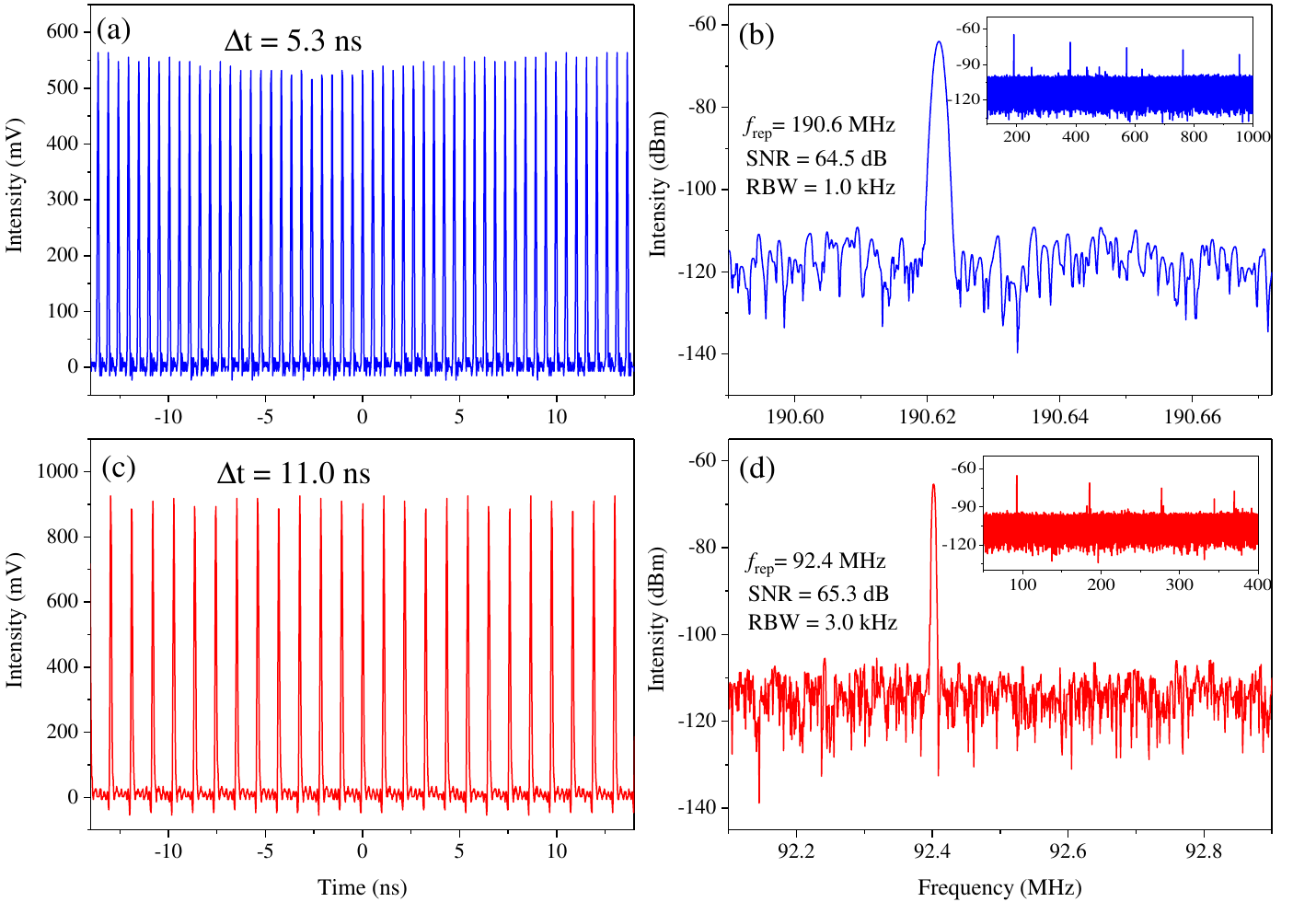}
	\caption{(a) Stable pulse train from the laser with 190.6~MHz repetition rate. (b) RF spectrum of the laser with 190.6~MHz repetition rate. Inset: the broad-span RF spectrum. (c) Stable pulse train from the laser with 92.4~MHz repetition rate. (d) RF spectrum of the laser with 92.4~MHz repetition rate. Inset: the broad-span RF spectrum.}
	\label{fig2}
\end{figure}

\begin{figure}[htbp]
	\centering
	\includegraphics[width=0.9\textwidth]{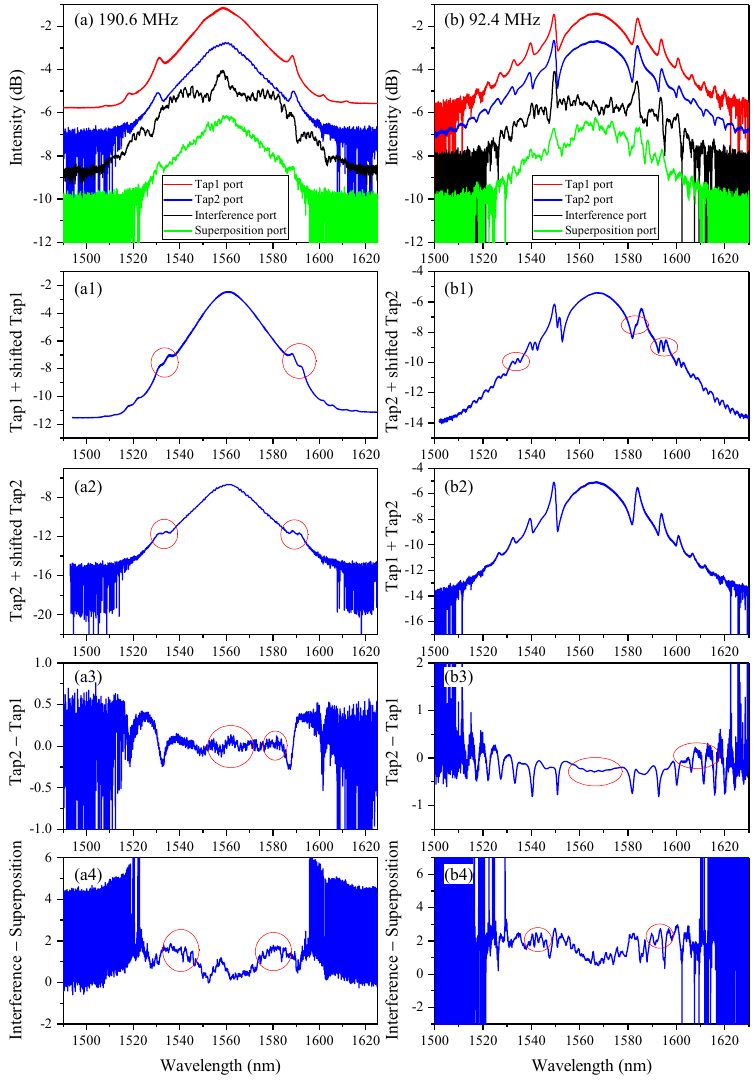}
	\caption{Optical spectra of the four output ports of the figure-9 lasers with repetition rates of (a) 190.6 MHz and (b) 92.4 MHz. The corresponding calculated spectra, based on the data from (a) and (b), are shown in (a1)–(a4) and (b1)–(b4), respectively. The spectral bases of the measured spectra are 1.35 dB apart.}
	\label{fig3}
\end{figure}

To investigate the impact of different nonlinear effects on the optical spectra, we reduced the repetition rate of a figure-9 laser from 190.6 MHz to approximately half (92.4 MHz) solely by increasing the passive fiber length between Col1 and Col4. The pulse train and radio frequency (RF) spectra were measured using an InGaAs biased detector (DET08CFC, Thorlabs), an oscilloscope (DSO-X 3054A, Agilent), and a spectrum analyzer (N9020A, Keysight); the results are shown in Fig.~\ref{fig2}. The data confirm that both lasers operate in a single-pulse regime. 
The corresponding optical spectra, measured using an optical spectrum analyzer (MS9710C, Anritsu), are shown in Fig.~\ref{fig3}(a) and Fig.~\ref{fig3}(b), for the 190.6 MHz and 92.4 MHz cases, respectively.
When the repetition rate was lowered from 190 MHz to 92.4 MHz, the spectrum of Tap1 port exhibited significant background noise while that of Tap2 port disappeared. Additionally, the spectral background noise increased at the interference port but remained essentially unchanged at the superposition port. We attribute this phenomenon to the optical power of the output pulses.

The pulses from the Tap1 and Tap2 ports both originate from the splitting of the same intracavity pulse. Therefore, their spectra can be regarded as essentially identical to that of the intracavity pulse. 
From the spectra of Tap1 port and Tap2 port in Fig.~\ref{fig3}(a) and Fig.~\ref{fig3}(b), it can be seen that the nonlinear effect growth brought by the increased passive fiber leads to an increase in both the number and amplitude of symmetric Kelly sidebands~\cite{kelly_characteristic_1992}. In contrast, the spectra from the superposition and interference ports differ considerably from the spectrum of the intracavity pulse, and exhibit severe distortion. Notably, the overall spectral profiles at the interference port are significantly distinct from those at other ports and remain stable.

To confirm whether nonlinear effects cause these spectral distortions and differences in spectral profiles, we performed numerical calculations on the experimental spectra from Fig.~\ref{fig3}(a) and Fig.~\ref{fig3}(b). For the interference port, light with different wavelengths propagates in clockwise and counterclockwise directions and interferes at PBS2. The resulting interference spectral intensity can be expressed as
\begin{equation}  
	I_{\mathrm{interf}}\left( \lambda \right) =\sum_i{I_{\mathrm{cw}}\left( \lambda _i \right) +I_{\mathrm{ccw}}\left( \lambda _i \right) +2\sqrt{I_{\mathrm{cw}}\left( \lambda _i \right) I_{\mathrm{ccw}}\left( \lambda _i \right)}\cos \left( \delta _i \right)}.  
\end{equation}  
Here, the third term represents the interference term and $\delta _i \geq \pi/2$. For convenience, we can set  
\begin{equation}  
	\sqrt{I_{\mathrm{ccw}}}=-\sqrt{I_{\mathrm{cw}}}\cos \left( \delta _i \right),  
\end{equation}  
yielding an interference spectral intensity independent of the nonlinear phase shift difference term $\cos \left( \delta _i \right)$:  
\begin{equation}\label{interf}  
	I_{\mathrm{interf}}\left( \lambda \right) =\sum_i{I_{\mathrm{cw}}\left( \lambda _i \right) -I_{\mathrm{ccw}}\left( \lambda _i \right)}.  
\end{equation}  
For the superposition port, since the polarization directions of the output light are mutually perpendicular, the superposition spectral intensity can be expressed as  
\begin{equation}\label{superpos}  
	I_{\mathrm{superpos}}\left( \lambda \right) =\sum_i{I_{\mathrm{cw}}\left( \lambda _i \right) +I_{\mathrm{ccw}}\left( \lambda _i \right)},  
\end{equation}  

Based on Equations (\ref{interf}) and (\ref{superpos}), the experimental spectra of the lasers are numerically processed to yield eight calculated spectra, as shown in Fig.~\ref{fig3}. It can be seen that some characteristics (marked with red circles) of these calculated spectra are consistent with the distorted spectrum characteristics of the superposition and interference ports. 
Specifically, since the spectral profiles of the clockwise and counterclockwise propagating light are nearly identical except for a minor shift, we calculated the superposition of the Tap2 spectrum with its 1 nm right-shifted counterpart. 
This procedure yielded multiple double-peak features, as shown in Fig.~\ref{fig3}(b1). These double-peak characteristics resemble the spectral profiles observed at the interference and superposition ports in Fig.~\ref{fig3}(b). 
Given that the intensities of the clockwise and counterclockwise pulses are comparable, the contributions of SPM and self-steepening effects to the spectral shift are negligible. Therefore, the introduced spectral shift is primarily attributed to the non-reciprocal nature of XPM, which drives the width and peak position of the Kelly sidebands between the Tap1 port and Tap2 port spectra in Fig.~\ref{fig3}(b) to differentiate.

In Fig.~\ref{fig3}(b2), the numerical superposition of the Tap2 port and Tap1 port spectra does not exhibit significant features. It is because the spectral shift of the counterclockwise transmitted pulses is small before experiencing the amplification.
Consequently, in Fig.~\ref{fig3}(a1) and Fig.~\ref{fig3}(a2), we only performed the superposition calculation of the Tap1 port and Tap2 port spectra with their own spectra right-shifted by 3 nm, respectively. These calculations likewise yielded characteristics similar to the spectra from the interference and superposition ports. The increased shift magnitude was necessary because the spectral structures generated by a smaller shift were not pronounced. Specifically, the calculation in Fig.~\ref{fig3}(a1) reproduced the significant intensity drop between 1530 nm and 1590 nm observed in the interference port spectrum of Fig.~\ref{fig3}(a). Meanwhile, the calculation in Fig.~\ref{fig3}(a2) successfully replicated the peak structure spanning from 1530 nm to 1590 nm on a flat substrate in the interference port spectrum of Fig.~\ref{fig3}(a).

In Fig.~\ref{fig3}(b3), an interference calculation between the Tap2 port and Tap1 port spectra resulted in a flattened central profile, which resembles the spectral feature observed at the interference port in Fig.~\ref{fig3}(b). This indicates that the flattening of the central spectrum between 1555–-1575 nm in Fig.~\ref{fig3}(b) is due to partial destructive interference between the counterclockwise and clockwise light, suggesting that this portion of the light lies within the reverse saturation absorption region of the NALM. Similarly, the small peak within the otherwise flattened central spectrum of the interference port in Fig.~\ref{fig3}(a) also originates from partial destructive interference. The key difference, however, is that the central spectral intensity of one beam is significantly higher than the other, leading to the formation of this small peak, as shown in Fig.~\ref{fig3}(a3).

\begin{figure}[htbp]
	\centering
	\includegraphics[width=0.8\textwidth]{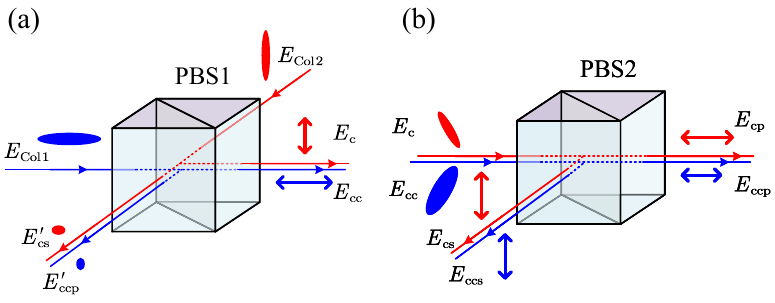}
	\includegraphics[width=0.6\textwidth]{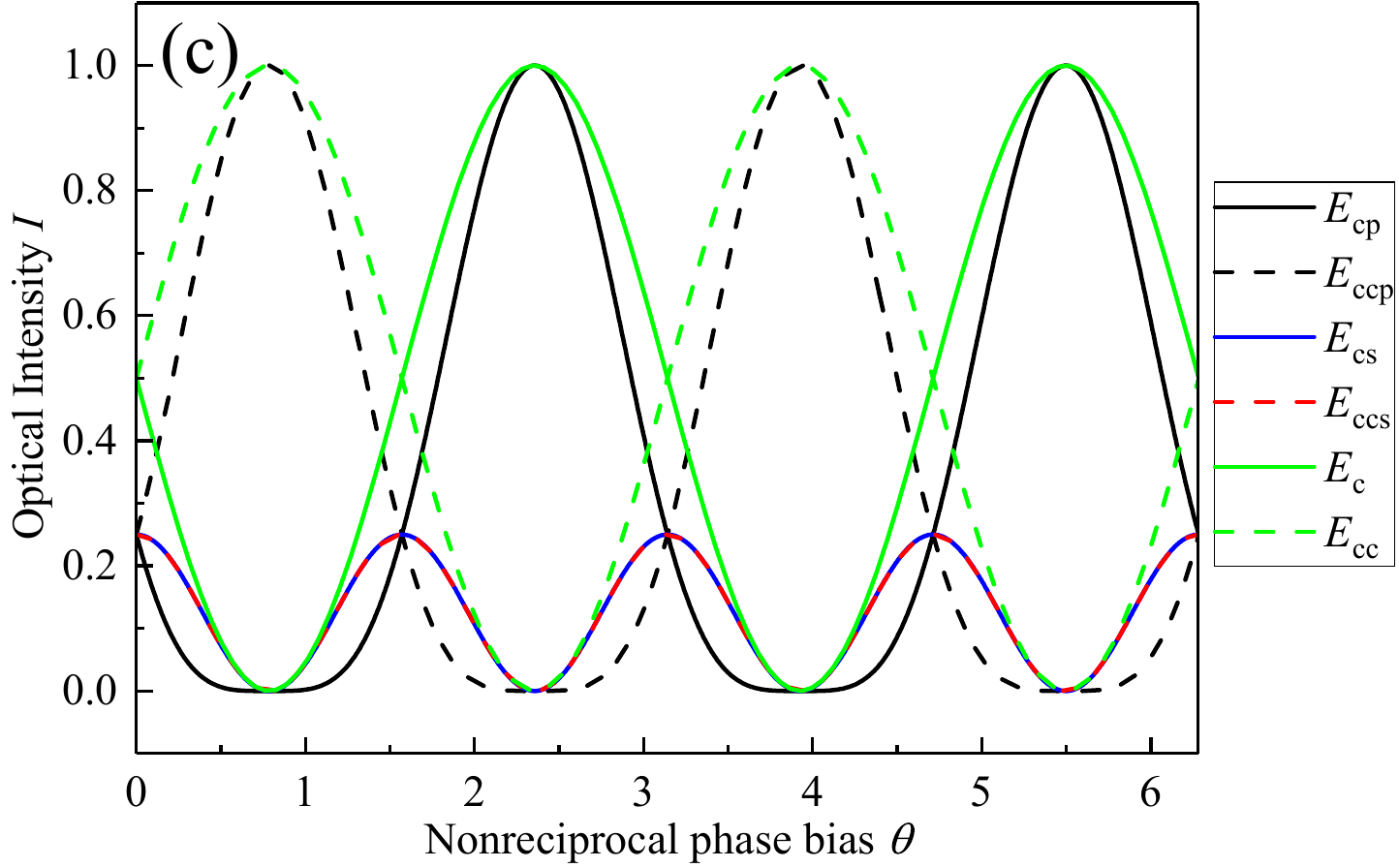}
	\caption{(a) Schematic diagram of beam splitting on PBS1. (b) Schematic diagram of beam splitting on PBS2. (c) The optical intensity of each component varies with nonreciprocal phase bias $\theta$.}
	\label{fig4}
\end{figure}

In Fig.~\ref{fig3}(a4) and Fig.~\ref{fig3}(b4), the spectra of the interference port and the superposition port were subtracted to distinguish the respective impacts of interference and superposition on the spectral profiles. It is evident that the characteristic features of the interference port in Fig.~\ref{fig3}(b) are essentially preserved, indicating that the spectral distortions in the interference port primarily originate from interference effects. The observed central dip suggests that the destructive interference effect in the superposition port spectrum is weak.
This occurs because, as illustrated in Fig.~\ref{fig4}(a), when the counterclockwise and clockwise beams are combined at PBS1 within the fiber loop, a portion of their energy leaks out of the cavity in the form of an orthogonal polarization state. 

Additionally, the output light contains a continuous-wave component. This conclusion can be drawn from comparing the average power across the four output ports in Table~\ref{tab1}, where the optical power for the superposition port was measured after the output light passed through a dichroic mirror (DMLP1180, Thorlabs). The $s$- and $p$-states of PBS1 are parallel to the slow axes of Col1 and Col2, respectively, resulting in a light leakage of less than 1\% at PBS1. Nevertheless, the optical power at the superposition port is more than twice that at the TCWDM with a 10\% splitting ratio. This indicates the simultaneous presence of elliptically polarized continuous-wave light generated by amplified spontaneous emission in the erbium-doped fiber, which leads to the superposition port having the highest output power. Due to its dual-polarization characteristics and significant continuous-wave component content, the superposition port is not suitable for practical applications.
\begin{table}[!t]
	\centering
	\caption{Optical powers from the four output ports of the figure-9 laser (190.6 MHz) under mode-locked (1300 mW pump) and continuous-wave (1240 mW pump) operations.}\label{tab1}
	\begin{tabular}{ccc}
		\hline
		Output port&Mode-locked state&Continuous-wave state\\
		\hline
		Superposition port& 41.3 mW & 33.5 mW\\
		Interference port& 22.8 mW & 66.1 mW\\
		Tap1 port & 2.6 mW & 1.7 mW \\
		Tap2 port & 14.1 mW & 12.3 mW \\
		\hline
	\end{tabular}
\end{table}

Since the saturable absorption effect in the figure-9 laser originates from the interference of two beams with a nonlinear phase shift difference, we can also provide a theoretical explanation from the perspective of interference theory for the causes of the spectral distortions and the substantial spectral differences between the intracavity pulse and the pulses at the external superposition and interference ports.

For convenience, we assume that horizontally polarized light transmitted from PBS2 to the left has a unit optical intensity and we ignore the beam splitting of TCWDM, the gain as well as the differences in bidirectional amplification within the fiber loop. The two beams of light, $E_\mathrm{c}$ and $E_\mathrm{cc}$, transmitted clockwise and counterclockwise in the fiber loop, return to PBS2 for beam splitting, and their identical components interfere with each other, as depicted in Fig.~\ref{fig4}(b). According to the Jones matrix of the nonreciprocal phase shifter~\cite{jones_new_1941,li_all-polarization-maintaining_2018}
\begin{equation}
	J_{\mathrm{NPS}} = \left[ \begin{matrix}
		\cos \left( \beta \pm \theta \right)&		\sin \left( \beta \pm \theta \right)\\
		-\sin \left( \beta \pm \theta \right)&		\cos \left( \beta \pm \theta \right)\\
	\end{matrix} \right] ,
\end{equation}
the corresponding components can be expressed as
\begin{align}
	E_{\mathrm{cs}}&=\cos \left(\beta - \theta \right)\cos \left(\beta + \theta \right) \exp \left( i\phi _{\mathrm{c}} \right),\\
	E_{\mathrm{ccs}}&=\sin \left(\beta - \theta \right)\sin \left(\beta + \theta \right) \exp \left( i\phi _{\mathrm{cc}} \right),\\
	E_{\mathrm{cp}}&=-\sin \left(\beta - \theta\right) \cos \left(\beta + \theta \right) \exp \left( i\phi _{\mathrm{c}} \right),\\
	E_{\mathrm{ccp}}&=\cos \left(\beta - \theta \right)\sin \left(\beta + \theta \right) \exp \left( i\phi _{\mathrm{cc}} \right).
\end{align}
Here, $\beta$ and $\pm \theta$ are the reciprocal and nonreciprocal phase bias, respectively. $\phi _\mathrm{c}$ and $\phi _\mathrm{cc}$ represent the nonlinear phase shifts accumulated by $E_\mathrm{c}$ and $E_\mathrm{cc}$, respectively. $E_\mathrm{cs}$ and $E_\mathrm{ccs}$ are the $s$-components reflected by PBS2, while $E_\mathrm{cp}$ and $E_\mathrm{ccp}$ are the $p$-components transmitted through PBS2. Fig.~\ref{fig4}(c) shows the optical intensity of each component varies with nonreciprocal phase bias $\theta$.
Since $\beta$ is the $\pi /4$ rotation angle of FR, it follows from the equations that $E_\mathrm{cs} = E_\mathrm{ccs}$, and the angle $\theta$ only affects the relative amplitudes of the intracavity components $E_\mathrm{cp}$ and $E_\mathrm{ccp}$. This precisely explains the significant spectral differences observed in Fig.~\ref{fig3} between the intracavity pulse and the pulses at the superposition and interference ports. 
Furthermore, under different pump powers, due to the difference of the actual gain in fiber loop, the amplitude difference between $E_\mathrm{cs}$ and $E_\mathrm{ccs}$ also varies, and their interference results in periodic peaks at the center of the spectrum~\cite{zhang_observation_2024}. 
When the pump power is at the minimum value required to sustain mode-locking, the intensity difference between the two pulse components at the interference port is small, resulting in better coherence of the output pulses.

As shown in Fig.~\ref{fig4}(b), the two light $E_\mathrm{c}$ and $E_\mathrm{cc}$ interfere at PBS2, the output light intensity is
\begin{equation}
		I_R=\left| E_{\mathrm{cs}}+E_{\mathrm{ccs}} \right|^2 =\cos ^4\theta +\sin ^4\theta +2 \sin ^2\theta \cos ^2\theta \cos \left( \Delta \phi \right) .
\end{equation}
Here, $\Delta \phi$ is the differential nonlinear phase shift. For the interference port, the $\Delta \phi$ of continuous-wave light is close to zero, so the light is almost entirely output. On the other hand, pulsed light experiences gain difference between forward and backward amplification, as well as noise in amplification of erbium-doped fiber, and this leads to changes in the output light intensity $I_R$. These factors cause the fluctuations in light intensity as a function of wavelength, resulting in small spectral spikes shown in Fig.~\ref{fig3}(a) and Fig.~\ref{fig3}(b). As the superposition port has less interference in its light beams compared to the interference port, there is smaller spectral spikes and spectral changes.

\begin{figure}[htbp]
	\centering
	\includegraphics[width=0.6\textwidth]{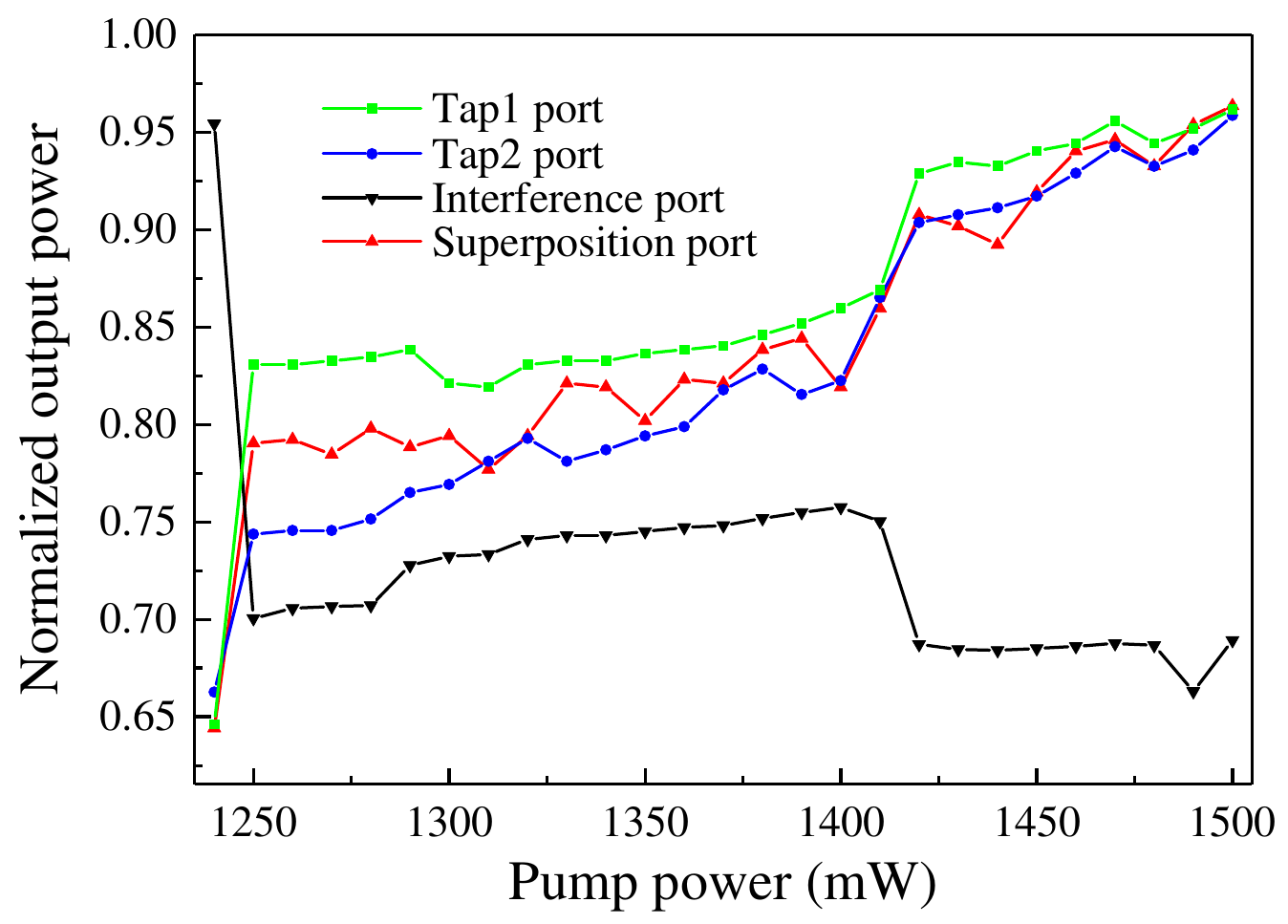}
	\caption{Changes in normalized output power of four output ports of the mode-locked laser as pump power gradually decreases.}
	\label{fig5}
\end{figure}

After achieving single-pulse mode-locking at the maximum available pump power of 1500 mW, the pump power was gradually decreased in steps of 10 mW. The resulting output power evolution for each port is shown in Fig.~\ref{fig5}. The output power values were normalized to facilitate a clearer comparison of their relative variations. While previous studies have included output power diagrams, they typically presented data from either a single port or multiple ports without normalization, making it difficult to assess the correlations between different outputs. As illustrated in Fig.~\ref{fig5}, the interference port exhibits an overall variation trend opposite to that of the other three ports, indicating that the output loss of the NALM actually increases as the pump power decreases.
In terms of curve smoothness, the Tap1 port and interference port curves are the smoothest, followed by that of Tap2 port, while the superposition port shows the least smooth trend. This behavior can be attributed to the fact that the light from the superposition port is reflected back into the fiber loop and eventually emitted from Tap1 port, whereas the outputs from the superposition and Tap2 ports are both amplified by the gain fiber.

When the pump power is reduced to approximately 1430 mW, the output power of the interference port increases abruptly, while the other three ports exhibit a simultaneous decrease. This is because, similar to the multi-pulse regime, there are multistability states in the single-pulse regime, and the sudden change of the output power is caused by a transition between different stability states\cite{komarov_multistability_2005,tang_mechanism_2005,zhang_observation_2024}.
Furthermore, when the pump power drops to 1240 mW, mode-locking can no longer be sustained. The corresponding output powers for the four ports are shown in Table~\ref{tab1}. A significant increase is observed in the interference port output, while the other three ports show decreases of a similar extent. This distinct behavior stems from the specific cavity configuration of the figure-9 laser, which leads to a strong contrast in the interference port output depending on the presence or absence of saturable absorption effects. This property can thus serve as an indicator for the mode-locking state of the laser.

\section{Conclusion}
In conclusion, we proposed a beam splitter TCWDM with little spectral impact, and utilize it to construct two all-PM figure-9 fiber lasers. The TCWDM enables the laser to output two pulses that are transmitted in the counterclockwise and clockwise directions of the fiber loop without superposition and interference. Through numerical calculations of the spectra obtained via the TCWDM, we identified the origin of several distinctive spectral features. These mechanisms are mainly related to the non-reciprocity of XPM, and the resulting features are consistent with the experimental spectral features of the pulses from the other two ports. 
Further theoretical investigation into the saturable absorption mechanism reveals that the observed spectral distortions are aggravated by spectral superposition and interference, rather than by the commonly believed nonlinear effects. Moreover, the stable existence of the significant spectral difference between the intracavity pulse and the output pulses can be explained by the interference conditions at PBS2: the intracavity $p$-components of the two beams always interfere with unequal intensity, whereas the output $s$-components typically interfere with equal intensity.
These results clarify the characteristics of each output port, provide new perspectives for simulating spectra that closely resemble experimental results, and can help interpret common experimental phenomena and deepen our understanding of spectral evolution and pulse dynamics. If a better beam-splitting scheme or more output ports are introduced, it may enable a more comprehensive characterization of the spectral evolution.

\begin{backmatter}
\bmsection{Funding}
This work was supported by National Natural Science Foundation of China under Grants 62105368, 62275268, and 62375284.


\bmsection{Disclosures}
The authors declare no conflicts of interest.

\end{backmatter}


\bibliography{Reference}

\end{document}